\documentclass[aps,twocolumn,prb]{revtex4}
\usepackage[scr=boondoxo,scrscaled=1.05]{mathalfa}
\usepackage{tikz}
\usetikzlibrary{calc,decorations.markings}
\usepackage{epsfig}
\usepackage{graphicx}
\usepackage{rotating}
\usepackage{amssymb}
\usepackage{amsmath}
\usepackage{float}   
\usepackage[active]{srcltx}
\usepackage{url}
\usepackage{hyperref}
\usepackage{centernot}
\usepackage{dsfont,bm}
\usepackage{xcolor}
\usepackage{pgfplots}
\usepackage{float}  
\usepackage[scr=boondoxo,scrscaled=1.05]{mathalfa}
\hypersetup{
	colorlinks   = true, 
	urlcolor     = blue, 
	linkcolor    = blue, 
	citecolor   = blue 
}
\usepackage{tikz}
\usetikzlibrary{calc,decorations.markings}

\DeclareMathOperator{\tr}{tr}
\usepackage{soul}

\colorlet{darkred}{red!55!black}
\colorlet{darkgreen}{green!25!black}
\graphicspath{{"/Users/donvil/OneDrive - University of Helsinki/Research/Calorimeter-Projects/Imperfect_Measurement/figures/"}}

\begin{document}

\title{Apparent Heating due to Imperfect Calorimetric Measurements}

\author{Brecht Donvil}
\affiliation{University of Helsinki, Department of Mathematics and Statistics
    P.O. Box 68 FIN-00014, Helsinki, Finland}
\email{brecht.donvil@helsinki.fi}
\author{Joachim Ankerhold}
\affiliation{Institute for Complex Quantum Systems and IQST, Ulm University - Albert-Einstein-Allee 11, D-89069 Ulm, Germany}
\email{joachim.ankerhold@uni-ulm.de}

\begin{abstract}
Performing imperfect or noisy measurements on a quantum system both impacts the measurement outcome and the state of the system after the measurement. In this paper we are concerned with imperfect calorimetric measurements. In calorimetric measurements one typically measures the energy of a thermal environment to extract information about the system. The measurement is imperfect in the sense that we simultaneously measure the energy of the calorimeter and an additional noise bath. Under weak coupling assumptions, we find that the presence of the noise bath manifests itself by modifying the jump rates of the reduced system dynamics. We study an example of a driven qubit interacting with resonant bosons calorimeter and find increasing the noise leads to a reduction in the power flowing from qubit to calorimeter and thus an apparent heating up of the calorimeter. 
\end{abstract}
\maketitle

\section{Introduction}
Measurements unavoidably come with noise. In quantum mechanics the effect 
of a noisy measurement is double, as it does not only influence the outcome but also the state of the system after the measurement.

Perfect continuous measurements lead to the quantum Zeno effect, where the system gets stuck in a state with a small probability to jump away \cite{Misra1977}. For imperfect measurements the dynamics are much richer. In 
case measurement times are sufficiently short and the distribution of measurement outcomes is much broader than the state vector, the dynamics of the measured system are described by a non-linear stochastic Liouville equation \cite{Belavkin1987}, see also \cite{Jacobs_2006} for a pedagogic introduction. Intermediate regimes regimes have been successfully studied with path-integral methods \cite{Mensky1993,Audretsch1997}.
The dynamics of a system weakly coupled to an environment under continuous perfect measurements can be described by quantum jump equations \cite{HuPa1984,BaBe1991,GaPaZo1992,DaCaMo1992,Car1993}, see also \cite{BrPe2020,WisemanBoek}. If the perfect measurement is disturbed by the presence of another bath and one considers a coarse grained time scale on which many jumps happen, the system dynamics undergo quantum state diffusion \cite{BrPe2020,WisemanBoek}.

In quantum calorimetric measurement schemes one continuously measures the 
energy, or temperature, of the environment in contact with a system of interest. These indirect measurements allow to indirectly extract information about the system. Quantum calorimetry has been used for example to detect cosmic x-rays \cite{Stahle1999} and in quantum circuit measurements \cite{Ronzani2018,Kokkoniemi2019,Senior2020}. Recent experiments have shown that quantum calorimeters form a promising tool for single microwave photon 
detection in quantum circuits  \cite{Karimi2020}. 

Earlier studies of experimental setups such as \cite{Ronzani2018} have modelled the calorimeter-system dynamics as coupled jump processes of the energy \cite{SuKu2016} or temperature of calorimeter \cite{KuMu16,DonvilCal,Donvil2019} and state of the system. The approaches of \cite{SuKu2016,KuMu16,DonvilCal,Donvil2019} differ from the usual quantum jump schemes\cite{HuPa1984,BaBe1991,GaPaZo1992,DaCaMo1992,Car1993} in that they explicitly update the state of the environment with the measured value of the energy or temperature. Concretely, this means that the jump rates depend on the measured state of the calorimeter. 
The previous coupled jump equations \cite{SuKu2016,KuMu16,DonvilCal,Donvil2019} all require that the calorimeter was under continuous {\em perfect} energy or temperature measurements. In the current work, we study quantum calorimetric measurement set-ups such as \cite{Ronzani2018} in the case of {\em imperfect} detection. 

The authors of \cite{Warszawski2002} developed quantum trajectories for realistic photon detection. They take into account noise on the measurement outcome and delay in obtaining it. Recently these ideas were applied to 
model single photon measurements in quantum circuit calorimetric measurements \cite{Karimi2020a}. Our approach differs from \cite{Warszawski2002,Karimi2020a} in the sense that we do not consider noise or delay on the outcome, i.e. imperfect information, but noise on the actual projective operator applied to the calorimeter during the measurement. 

We consider a simple model for the imperfect measurement by introducing another bath, the noise bath, that does not interact with calorimeter or system. The only influence of the noise bath is that its energy is measured  simultaneously with the calorimeter energy, instead of just the calorimeter energy. This is similar to the setup where quantum state diffusion applies \cite{BrPe2020,WisemanBoek}, although we are not interested in coarse graining time. Under the presence of the noise bath we derive a hybrid master equation and corresponding coupled jump process for the measured energy and system state. Comparing out result to the dynamics 
described in \cite{SuKu2016,KuMu16,DonvilCal,Donvil2019}, we find that the coupled jump equations have similar structures but the jump rates are modified due to the presence of the noise bath. Note that our approach is not limited to energy measurements but applies to any macroscopic property of the finite size environment such as for example magnetism.

The paper is structured as follows: in Section \ref{sec:model} we introduce our model, a qubit interacting with the calorimeter and additional noise bath, and how we model the imperfect measurements. In Section \ref{sec:hybrid} we derive the hybrid master equation for the measured calorimeter-noise bath energy and the qubit state. We also give the corresponding energy-qubit state jump process. 
We consider the specific example of a calorimeter consisting of resonant bosons in Section \ref{sec:example}. We study the effect of the noise bath on the rates and on the power flowing to the calorimeter. In Section \ref{sec:exp} we discuss experimental setups in which the effects of imperfect measurements could play a role. Finally, in Section \ref{sec:discussion} we discuss the results and provide an outlook.

\section{Model}\label{sec:model}
\begin{figure}
\centering
\includegraphics[scale=1]{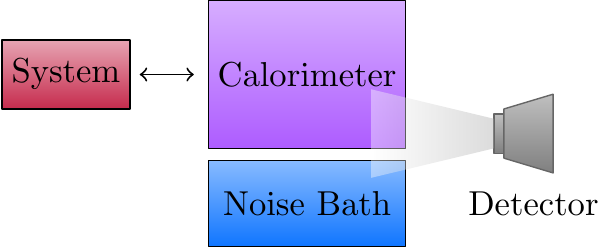}
\caption{Schematic representation of the system-calorimeter-noise bath setup. The system and calorimeter interact while the energy of the calorimeter and noise bath is simultaneously measured.}
\label{fig:setup}
\end{figure}

We consider a system interacting with a calorimeter  (see fig.~\ref{fig:setup}) modelled as a finite size environment. A detector permanently measures the combined energy of this calorimeter and an additional noise bath 
which, however, is not directly coupled to system and calorimeter. Here we introduce the model and below, in Sec.~\ref{sec:hybrid}, we derive a joint master equation for the state of the system and the energy monitored by the detector. 

While the framework developed below can be applied to any finite dimensional system interacting with arbitrary reservoirs, for the sake of simplicity and being explicit, we here consider a  paradigmatic set-up: a qubit linearly coupled a bosonic calorimeter. The total Hamiltonian then consist of
\begin{equation}
H_Q=\omega \sigma_+\sigma_-
\label{eq:qubit}
\end{equation}
for the system (qubit),
\begin{equation}
H_B=\sum_k\omega_k b_k^\dagger b_k\, 
\end{equation}
for the boson reservoir, i.e. the calorimeter, and
\begin{equation}
H_I=\sum_kg_k(\sigma_+b_k+\sigma_-b^\dagger_k)
\label{eq:interactionH}
\end{equation}
for the interaction between system and calorimeter. 
Finally, we write the noise bath Hamiltonian as 
\begin{equation}
H_N=\sum_E E \operatorname{P}^N_E,
\end{equation}
where orthogonal projectors $\operatorname{P}^N_E$ project on states of energy $E$ of the noise bath.
This way, together with the  corresponding operator for the boson reservoir  $\operatorname{P}_E$, we can now introduce  the projector $\mathcal{P}_E$ on the joint reservoir and noise bath energy $E$, i.e.,  
\begin{equation}
\mathcal{P}_E=\sum_{E'}\operatorname{P}_{E'}\otimes \operatorname{P}^N_{E-E'}
\end{equation}

In our derivation of the energy-qubit state master equation below, we follow the usual weak coupling treatment and assume that the qubit-calorimeter-noise bath initial state is of the form
\begin{equation}\label{eq:initial_state}
\rho_0=\sum_E\rho(E,0)\otimes \mathcal{P}_E
\end{equation}
with the  partially reduced density $\rho(E,0)$ of the qubit at fixed energy $E$ of the total environment.
Accordingly, taking the trace $\tr(\mathcal{P}_E\rho_0)=\tr_Q{\rho(E,0)}$, we observe that the trace of the unnormalised qubit state $\rho(E,0)$ 
is the probability to measure the environmental energy $E$ at the initial 
time.
Eliminating the noise reservoir, leads us to the qubit-calorimeter state
\begin{equation}
\tr_N \rho_0=\sum_E\rho(E,0)\otimes \sum_{E'} \operatorname{P}_{E'} \tr_N\{\operatorname{P}^N_{E-E'}\}\, .
\end{equation}
The latter distribution captures the energy partition between calorimeter 
and the noise bath which for the sake of simplicity is taken to be of the 
form
\begin{equation}\label{eq:guassian-noise}
\tr_N\{\operatorname{P}^N_{E-E'}\}\propto e^{-(E-E')^2/k}
\end{equation}
with a tunable parameter $k$. However, any other model originating, for example, from a more microscopic description for a specific experimental set-up is applicable as well.

\section{Derivation of the Hybrid Master Equation}\label{sec:hybrid}

We are now in a position to derive a qubit-energy hybrid master equation for the partially reduced density $\rho(E,t)$ by generalizing the usual steps of a weak coupling Born-Markov treatment \cite{BrPe2020}.  An alternative route in terms of the Nakajima-Zwanzig projection operator technique \cite{Nakajima1958,Zwanzig1960} is presented in Appendix \ref{sec:NJ}, see also \cite{Steinigeweg2007,Mallayyaz2019} for similar applications. 

We start with the interaction-picture dynamics of the qubit-calorimeter-noise bath as described by the Liouville-von Neumann equation
\begin{equation}\label{eq:liouville-von-neumann}
\frac{d}{dt}\tilde{\rho}(t)=-i[\tilde{H}_I(t),\tilde{\rho}(t)]
\end{equation}
with the interaction picture Hamiltonian defined as
\begin{align}\label{eq:interaction-hamiltonian}
\tilde{H}_I(t)=&e^{-i(H_Q+H_B)t}H_Ie^{i(H_Q+H_B)t}\nonumber\\
=&\sum_kg_k\left[e^{i(\omega-\omega_k)t}\sigma_+b_k+e^{-i(\omega-\omega_k)t}\sigma_-b^\dagger_k\right]
\end{align}
and $\tilde{\rho}$ the corresponding interaction picture state operator. 

\paragraph*{Born-Markov approximation.} 
Integrating equation \eqref{eq:liouville-von-neumann} and plugging it back into itself gives
\begin{equation}\label{eq:liouville-von-neumann2}
\frac{d}{dt}\tilde{\rho}(t)=-i[\tilde{H}_I(t),\rho_0]-\int_0^t ds\, [\tilde{H}_I(t),[\tilde{H}_I(s),\tilde{\rho}(s)]].
\end{equation}
To trace out the calorimeter and noise bath degrees of freedom, we proceed by applying a modified version of the \textit{Born approximation}. Since the calorimeter-noise bath energy is continuously measured, we assume that at any time $t$ the qubit-reservoir-noise bath state $\tilde{\rho}(t)$ has the same structure as the initial state \eqref{eq:initial_state}, i.e. 
\begin{equation}\label{eq:born-assumption}
\tilde{\rho}(t)\approx\sum_E\tilde{\rho}(E,t)\otimes \mathcal{P}_E.
\end{equation}
Additionally, we perform the Markov approximation by replacing $\tilde{\rho}(s)$ in the integral in equation \eqref{eq:liouville-von-neumann2} by $\tilde{\rho}(t)$. Taking the partial trace $\tr_{C+N}(\mathcal{P}_E\,\, .\,\,)$ on both sides of \eqref{eq:liouville-von-neumann2} gives
\begin{align}\label{eq:liouville-von-neumann3}
&\frac{d}{dt}\tilde{\rho}(E,t)=-i\tr_{C+N}(\mathcal{P}_E[\tilde{H}_I(t),\rho_0])\nonumber\\
&\quad-\int_0^t ds\, \tr_{C+N}(\mathcal{P}_E[\tilde{H}_I(t),[\tilde{H}_I(s),\sum_{E'}\tilde{\rho}(E',t)\otimes \mathcal{P}_{E'}]].
\end{align}
We observe that traces $\tr(\mathcal{P}_E b_k\mathcal{P}_{E'})=\tr(\mathcal{P}_E b^\dagger_k\mathcal{P}_{E'})=0$, which allows us to conclude that the first term on the right hand side of the first line of \eqref{eq:liouville-von-neumann3} is zero. Next, the  change of variables $s\rightarrow t-s$ is performed in the remaining integral  and it is assumed that 
the calorimeter-noise correlation functions decay over a time scale $\tau_B$ much smaller than the qubit relaxation time $\tau_R$. Under said assumption we are permitted to let the upper limit of the integral in \eqref{eq:liouville-von-neumann3} go to infinity to arrive at the master equation
\begin{align}\label{eq:liouville-von-neumann4}
&\frac{d}{dt}\tilde{\rho}(E,t)=\nonumber\\&\quad-\int_0^\infty ds\, \tr_{C+N}(\mathcal{P}_E\tilde{H}_I(t),[\tilde{H}_I(t-s),\sum_{E'}\tilde{\rho}(E',t)\otimes \mathcal{P}_{E'}]].
\end{align}

\paragraph*{Master Equation.} 
From equation \eqref{eq:liouville-von-neumann4} we proceed similar to the usual derivation of the Lindblad equation \cite{BrPe2020}. Evaluating the time integral, reordering terms and dropping the Lamb shift, we arrive in the Schr\"{o}dinger picture at the hybrid master equation
\begin{align}\label{eq:master}
&\frac{d}{dt}\rho(E, t)=-i[H_Q,\rho(t)]\nonumber\\&+\Gamma_{\uparrow}(E+\omega) \sigma_+\rho(E+\omega,t)\sigma_- -\frac{\Gamma_\uparrow(E) }{2}\{\sigma_-\sigma_+,\rho(E,t)\}\nonumber\\
&+\Gamma_\downarrow(E-\omega)\sigma_-\rho(E-\omega,t)\sigma_+-\frac{\Gamma_\downarrow(E)}{2}\{\sigma_+\sigma_-,\rho(E,t)\}.
\end{align}
The transition rates at fixed environmental energy $E$ are given by 
\begin{subequations}\label{eq:rates1}
\begin{align}
\Gamma_\uparrow(E) =& \kappa(\omega)\frac{\tr_{C+N}\{b^\dagger_\omega b_\omega \mathcal{P}_{E}\}}{\tr_{C+N}\{\mathcal{P}_{E}\}}\\
\Gamma_\downarrow(E) =&\kappa(\omega)\frac{\tr_{C+N}\{b_\omega b^\dagger_\omega \mathcal{P}_{E}\}}{\tr_{C+N}\{\mathcal{P}_{E}\}},
\end{align}
\end{subequations}
with  $b_\omega$ denoting the ladder operator of the reservoir mode resonant to the qubit frequency $\omega$ with coupling strength $\kappa(\omega)\propto g_\omega^2$ [see (\ref{eq:interactionH})].
Exploiting our assumption on the noise bath traces \eqref{eq:guassian-noise}, we obtain the more explicit expressions
\begin{subequations}\label{eq:rates}
\begin{align}
\Gamma_\uparrow(E) =&\kappa(\omega)\frac{\sum_{E'}\tr_{C}\{b^\dagger_\omega b_\omega \operatorname{P}_{E'}\} e^{-(E-E')^2/k} }{\sum_{E'}\tr_{C} \{\operatorname{P}_{E'}\} e^{-(E-E')^2/k} }\\
\Gamma_\downarrow(E) =&\kappa(\omega) \frac{\sum_{E'}\tr_{C}\{b_\omega b^\dagger_\omega \operatorname{P}_{E'}\} e^{-(E-E')^2/k} }{\sum_{E'}\tr_{C} \{\operatorname{P}_{E'}\} e^{-(E-E')^2/k} }.
\end{align}
\end{subequations}
These results reduce to their perfect measurement counterparts by considering $k\downarrow 0$ so that in \eqref{eq:rates}  $\mathcal{P}_{E}$ is replaced by the calorimeter energy projector $\operatorname{P}_{E}$ and in the sums over energies only the contribution with $E=E'$ survives \cite{SuKu2016}. Furthermore, for any finite $k$ and sufficiently large reservoirs we expect the size of the energy eigenspaces to increase with the energy. 
This implies $\tr_{C}\{\operatorname{P}_{E'}\},\,\tr_{C}\{b^\dagger_\omega b_\omega \operatorname{P}_{E'}\}$ and $\tr_{C}\{b_\omega b^\dagger_\omega \operatorname{P}_{E'}\}$ to grow with $E'$ and thus, as $k$ increases, 
the rates primarily feel increased contributions from higher energies. This in turn may appear as an effective heating up of the environment as, by way of example, will be qualitatively confirmed in Section \ref{sec:example}.

\paragraph*{Stochastic Evolution.}
The hybrid master equation (\ref{eq:master}) can be mapped onto an equivalent stochastic state vector dynamics according to 
\begin{equation}
\rho(E,t)=\mathsf{E}(|\psi(t)\rangle\langle\psi(t)|\delta(E-E(t))) \, ,\end{equation}
where $\mathsf{E}(\cdot)$ denotes a proper average over noise realizations. Accordingly, one finds the following set of coupled stochastic differential equations for the state of the qubit $\psi$ and the measured energy 
$E$ of the calorimeter-noise bath 
\begin{equation}\label{eq:stochastic-schrodinger}
\begin{cases}
d\psi(t)=&-iG(\psi(t))\,dt + \left(\frac{\sigma_-\psi(t)}{\|\sigma_-\psi(t)\|}-\psi(t)\right)dN_\downarrow\\&+ \left(\frac{\sigma_+\psi(t)}{\|\sigma_+\psi(t)\|}-\psi(t)\right)dN_\uparrow\\
dE(t)=&\omega(dN_\downarrow-dN_\uparrow)\, .
\end{cases}
\end{equation}
Here, the continuous evolution of the state vector is given by 
\begin{align*}
-iG(\psi)=&-i\omega\sigma_+\sigma_--\frac{1}{2}\Gamma_\downarrow(E(t))(\sigma_+\sigma_--\|\sigma_-\psi\|^2)\psi\nonumber\\&-\frac{1}{2}\Gamma_\uparrow(E(t))(\sigma_-\sigma_+-\|\sigma_+\psi\|^2)\psi
\end{align*}
and $N_\uparrow$, $N_\downarrow$ are Poisson processes with increments obeying 
\begin{align*}
\mathsf{E}(dN_\downarrow|\psi,E)=&\Gamma_\downarrow(E) \|\sigma_-\psi\|^2 dt\\
\mathsf{E}(dN_\uparrow|\psi,E)=&\Gamma_\uparrow(E) \|\sigma_+\psi\|^2 dt\, .
\end{align*}

\section{Example: Calorimeter of Resonant Oscillators}\label{sec:example}
\begin{figure}
\centering
\includegraphics[scale=1]{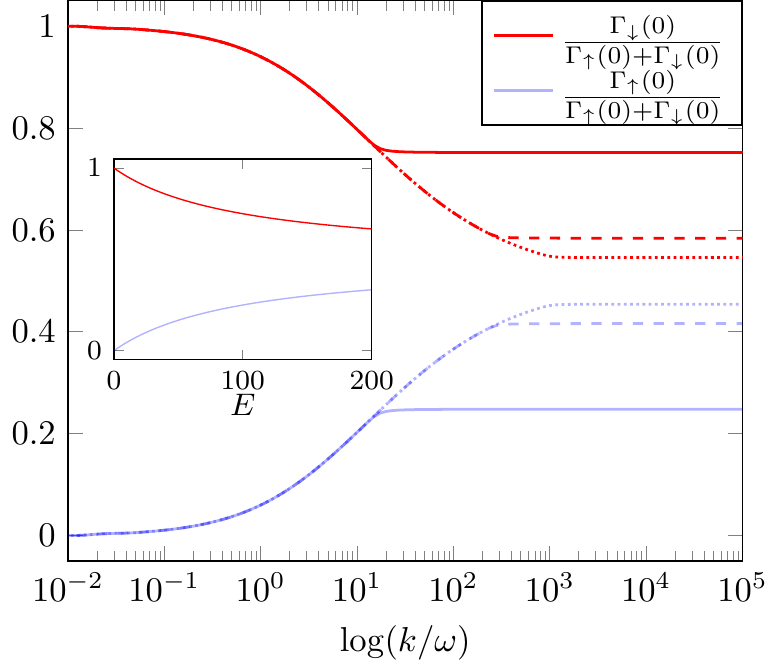}
\caption{Dependency of the rates \eqref{eq:rates} on $k$ for different values of the cut-off $N_C$ of the noise bath \eqref{eq:noise-bath-explicit}:  $N_C=100$ (solid), $N_C=500$ (dashed), $N_C=1000$ (dotted) for $E=0$. The inset shows the energy dependence of the rates in the case of perfect measurements (i.e. $k\rightarrow 0$).}
\label{fig:rates}
\end{figure}

The central ingredients for the hybrid master equation (\ref{eq:master}) are the rates (\ref{eq:rates}). By way of example and drawing on \cite{SuKu2016}, we study a system of $N$ resonant harmonic oscillators coupled to a qubit. In this case we are able to explicitly compute all traces required to compute the rates \eqref{eq:rates}. The number of micro-states of 
the calorimeter at a given energy $E=n \omega$, i.e. the size of the calorimeter energy eigenspaces,  is
\begin{equation}
\tr_C\{\operatorname{P}_{n\omega}\}={n+N-1 \choose N-1} 
\end{equation}
so that
\begin{subequations}
\begin{align}
&\tr_C\{a^\dagger_\omega a_\omega \operatorname{P}_{n\omega}\}=n {n+N-1 
\choose N-1}\\
&\tr_C\{a_\omega a^\dagger_\omega \operatorname{P}_{n\omega}\}=(n+N) {n+N-1 \choose N-1}\, .
\end{align}
\end{subequations}
Note that the above traces indeed grow with the value of $n$.
For the noise bath, we take 
\begin{equation}\label{eq:noise-bath-explicit}
H_N=\sum_{n=-N_C}^{N_C} P^N_{n\omega},\quad \tr_N\{P^N_{n\omega}\}\propto e^{-n^2 \omega^2/k}.
\end{equation}
The noise bath thus has two free parameters: $k$ and the cut-off $N_C$.
Figure \ref{fig:rates} shows the dependence of the rates on these parameters. As expected, we observe that the rate increase with $k$ and eventually plateau at a value which depends on $N_C$. The inset displays the energy dependence of the rates in the case of perfect measurements (i.e. $k\rightarrow 0$) which reveals that increasing $k$ has a similar effect as increasing $E$. Hence, in an imperfect detection considered here, the temperature of the calorimeter  appears to be elevated compared to the measured value of the energy.

\begin{figure}
\centering
\includegraphics[scale=1]{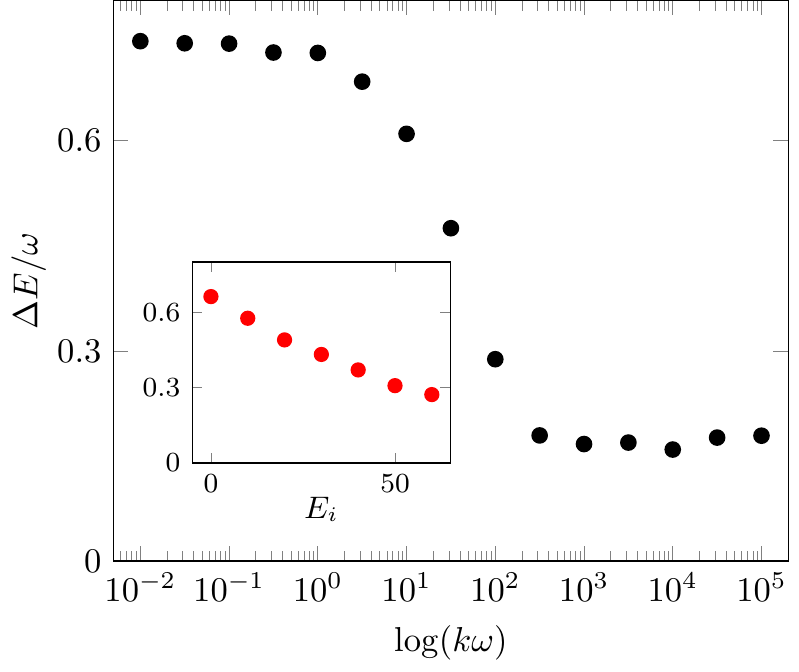}
\caption{Average change in measured energy of  the calorimeter after 5 periods of driving, the initial measured energy is $E_i=0$ and the qubit is initially in the ground state. The inset plot shows the measured power 
in the case of perfect measurements as a function the initial energy $E$. 
The parameters are $\kappa(\omega)=0.001\omega$, $\lambda= 0.05\omega$ and $N_C=100$.}
\label{fig:power-to-cal}
\end{figure}

\paragraph*{Energy flow to Calorimeter.}
In a next step, we explore how an imperfect measurement influences the detection of the energy flow towards the calorimeter when the qubit is driven by an external periodic signal, a generic situation  for quantum thermodynamical set-ups. For this purpose, a driving term is added to the qubit Hamiltonian (\ref{eq:qubit}), i.e.
\begin{equation}
H_Q(t)=\omega \sigma_+\sigma_-+\lambda(e^{i\omega t}\sigma_+e^{-i\omega 
t}\sigma_-)\, .
\end{equation}
The coupled jump equations \eqref{eq:stochastic-schrodinger} are simulated (time step $dt=0.03$) and the energy flow to the calorimeter is extracted  as a function of the energy uncertainty parameter $k$.  Figure \ref{fig:power-to-cal} reports the average change in the measured energy $\Delta E= E_f-E_i$ over 5 periods while the inset figure depicts the same quantity in the perfect measurement case as a function of the initial energy $E_i$ of the calorimeter. Obviously, in the latter case for increasing initial energy the average energy flowing to the calorimeter decreases. 
Imperfect measurements exhibit a similar behaviour for increasing $k$ which implies that effectively a less perfect measurement shows the same behavior as a hotter calorimeter.

\begin{figure}
\centering
\includegraphics[scale=1]{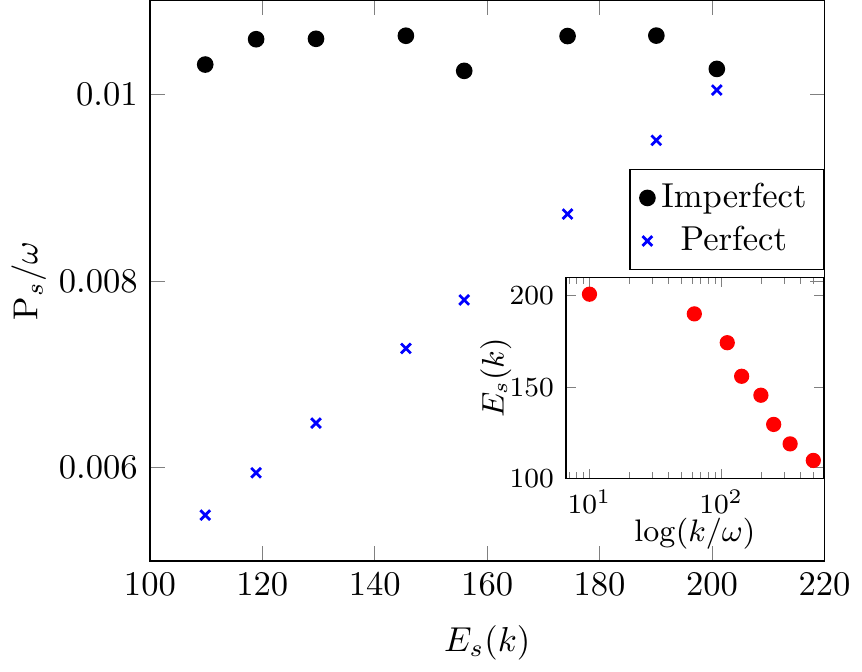}
\caption{Power at steady state as a function of the measured energy at the different values of $k$. The inset shows the measured energies as a function of $k$. The (black) dots show the estimate for the power at steady state \eqref{eq:power_at_ss} taking the value of $k$ into account, the (blue) crosses show the estimate without taking the imperfect measurement into account. The parameters are given in the caption of Figure \ref{fig:power-to-cal} and $\gamma=0.0005 \omega$.}
\label{fig:steady_state}
\end{figure}

\paragraph*{Power at Steady State.}
Finally, we address the power flowing through the qubit in steady state which in calorimetric measurements is  a viable tool to retrieve qubit properties
without `touching' the qubit directly \cite{Ronzani2018,Senior2020}. To study the steady state properties, we add the loss term $-\omega dN_{loss}$ to the energy jump process \eqref{eq:stochastic-schrodinger}, where
$N_{loss}$ is a Poisson process with rate
\begin{equation}
\mathsf{E}(dN_{loss})=\gamma \langle E \rangle dt\, 
\end{equation}
and $\langle E \rangle$ is the expected energy of the calorimeter when the energy $E$ was measured
\begin{equation}
\langle E \rangle=\frac{\sum_{E'}\tr_{C} E' \tr_{C} (\operatorname{P}_{E'}) e^{-(E-E')^2/k} }{\sum_{E'}\tr_{C} (\operatorname{P}_{E'})e^{-(E-E')^2/k} }\, .
\end{equation}
When the power from the qubit to the calorimeter and the loss term balance each other out, the calorimeter-noise bath reaches a steady state. We call the average energy at steady state $E_s$ and approximate the power flowing through the calorimeter by
\begin{equation}\label{eq:power_at_ss}
P_s=\gamma \langle E_s\rangle.
\end{equation}
Figure \ref{fig:steady_state} shows the power at steady state $P_s$ as a function of the measured average energy $E_s$ for different values of $k$. The inset shows the measured steady state energies $E_s$ as a function of $k$. As a comparison the (blue) crosses display the power at steady state when not taking the measurement error into account. In this case our estimate is then $P'_s=\gamma E_s$. We observe that in the imperfect situation the actual power flowing through the calorimeter is nearly constant. This is not surprising as the qubit driving remains constant when varying $k$. However, when not taking the measurement errors into account, 
a significant underestimation of the power can occur.

\section{Experimental Realisation}\label{sec:exp}

To put the above theoretical results into the context of current experimental activities, we discuss two different  experimental realizations, by way of example.

First, we think about the class of set-ups recently developed by Pekola and co-workers\cite{Ronzani2018,Senior2020,Karimi2020} also discussed in the introduction.  They consider a dedicated system, for example a transmon qubit, which is in thermal contact with a small metallic island (resistor). The Fermi gas of the latter acts as a calorimeter, the energy/temperature variations of which are monitored via rf-thermometry \cite{ScYu2003,GaVi2015}. The Fermi gas contains a sufficiently large number of electrons (typically on the order of about $10^8$) with sufficiently fast internal relaxation time compared to the electron-phonon coupling to qualify for a heat bath. Any residual on-chip noise source may feed energy in the detection lines and cannot be as clearly be identified as, for example, amplifier noise. 

A second example, refers to a Josephson Photonics set-up, where a dc-voltage biased Josephson junction is placed in series with a resonant mode of frequency $\omega_0$ (LC-oscillator, cavity) \cite{Hofheinz2011,Westig2017,Rolland2019}. For voltages below the superconducting gap, Cooper pairs can only be transferred 
if the excess energy $2eV$ can be deposited as electromagnetic excitation 
$\hbar\omega_0$ in the cavity. Due to the finite photon lifetime, photon radiation from the cavity can be then detected. These and related set-ups 
have recently be shown to operate as relatively bright sources for quantum microwaves. It turns out that the main source of `noise' is voltage noise at the junction, typically in the low frequency regime. Even spurious additional `resonances' may appear due to low frequency modes that are often unavoidable by circuit design. In fact, by slightly de-tuning the voltage from the resonance condition allows to effectively cool these resonant modes and thus to reduce the impact of voltage noise. A proper description  requires to extend master equations applied so far \cite{Gramich2013} to include also this additional harmonic low frequency (classical) `heat bath', for example, to explore quantum thermodynamics. The results presented here,  may provide a proper powerful theoretical framework.

\section{Discussion}\label{sec:discussion}
We analysed the role noise in calorimetric measurements by introducing a simple model for imperfect calorimetric measurements. We introduced a noise bath in addition to the system and the calorimeter. Only the last two interact, the role of the noise bath is that its energy is measured simultaneously as the calorimeter and as such disturbs the measurement.

Under weak coupling assumptions we derived a hybrid master equation for the system state and the measured energy \eqref{eq:master} and the corresponding stochastic evolution \eqref{eq:stochastic-schrodinger}. Remarkably, the only change compared to the perfect measurement case is the values of the jump rates. The rates average contributions from different energies which leads to an apparent heating of the calorimeter. We study a simple example of a driven qubit in contact with a reservoir of resonant bosons to study the qualitative behaviour of our model for imperfect measurements. As the measurement imperfection increases, the observed energy transfer from the qubit to the calorimeter decreases. 

To study the effect of noise in steady state measurements such as in \cite{Ronzani2018} to extract the power flowing from the qubit to the calorimeter, we add a loss term to the calorimeter energy process. We find that for growing imperfection of the measurement, not taking the imperfection into account leads to a significant underestimation of the power.

The model for imperfect measurements we presented can be applied to general systems and reservoirs. Further studies could include more complex systems, e.g. multiple qubits, and fermion reservoirs. Another potential avenue is to derive the noise bath from a microscopic description of the noise in an experimental setup. Changes in the noise bath distribution \eqref{eq:guassian-noise} could have a significant impact on the measurement outcomes.
\section{Acknowledgements}
We thank Paolo Muratore-Ginanneschi, Bayan Karimi, Jukka Pekola and Kalle 
Koskinen for useful discussions. B.D. acknowledges support from the AtMath collaboration at the University of Helsinki. This work has been supported by IQST, the Zeiss Foundation, and the German Science Foundation (DFG) under AN336/12-1 (For2724).

\appendix
\section{Nakajima-Zwanzig projection operator technique}\label{sec:NJ}
We define a projector $\operatorname{P}$, which acts on a qubit-calorimeter-noise bath state $\rho$ as
\begin{equation}
\operatorname{P}\rho=\sum_E \tr_{C+N}(\mathcal{P}_E\rho)\otimes\frac{\mathcal{P}_E}{\tr_{C+N}(\mathcal{P}_E)}.
\end{equation}
It is straightforward to check that $\operatorname{P}$ is indeed a projector, i.e. $\operatorname{P}^2\rho=\operatorname{P}\rho$. 

We introduce the orthogonal projector $\operatorname{Q}=\operatorname{1}-\operatorname{P}$, such that $\operatorname{Q}\operatorname{P}=\operatorname{P}\operatorname{Q}=0$ and define the hybrid state\begin{equation}
\rho(E,t)\equiv\tr(\mathcal{P}_E\operatorname{P}\rho)=\tr_{C+N}(\mathcal{P}_E\rho).
\end{equation} 
In order to apply the Nakajima-Zwanzig projection operator technique, see 
for example \cite{Rivas}, we require three hypotheses to be satisfied.
\paragraph*{Hypothesis 1:} The initial state \eqref{eq:initial_state} satisfies $\operatorname{P} \rho_0=\rho_0$. 
\paragraph*{Hypothesis 2:}
For any qubit-calorimeter-noise bath state $\rho$ we have that 
\begin{equation}
[H_C+H_N , \operatorname{P}\rho]=0.
\end{equation}
\paragraph*{Hypothesis 3:} 
For any qubit-calorimeter-noise bath state $\rho$ we have that 
\begin{equation}
\tr_{C+N}([H_I,\operatorname{P}\rho])=0.
\end{equation}
A direct computation shows that the above three hypotheses are satisfied for our model and choice of $\operatorname{P}$. Going though the Nakajima-Zwanzig projection operator technique, we find that $\rho(E,t)=\tr(\mathcal{P}_E\operatorname{P}\rho)$ satisfies a closed differential equation\begin{equation}\label{eq:nj-master}
\frac{d}{dt}\rho(E,t)=-\int_0^t ds \tr\left(\mathcal{P}_E [H_I(t),\operatorname{G}(t,s)[H_I(s),\operatorname{P}\rho]]\right)
\end{equation}
where $\operatorname{G}$ satisfies the differential equation
\begin{equation}
\frac{d}{dt}\operatorname{G}(t,s)=-i\operatorname{Q}[H_I(t),\operatorname{G}(t,s)].
\end{equation}
Up to lowest order in the coupling strength  $\operatorname{G}(t,s)\approx\operatorname{1}$, such that \eqref{eq:nj-master} equals \eqref{eq:liouville-von-neumann3}  up to second order in the coupling.

\bibliography{lit}
\end{document}